\documentclass[12pt]{article}
\usepackage{graphics}
\usepackage{amssymb}
\usepackage{graphics}
\usepackage{amssymb}
\usepackage[driver]{graphicx}

\begin{document}

\parskip=2pt
\parindent=7mm
\renewcommand{\baselinestretch}{1.}
\renewcommand{\theequation}{\arabic{section}.\arabic{equation}}

\newcommand{\be}{\begin{equation}}
\newcommand{\ee}{\end{equation}}
\newcommand{\ba}{\begin{eqnarray}}
\newcommand{\ea}{\end{eqnarray}}
\newcommand{\bd}{\begin{description}}
\newcommand{\ed}{\end{description}}
\newcommand{\pd}{\partial}
\newcommand{\e}{{\rm e}}
\newcommand{\D}{{\rm d}}
\newcommand{\ageq}{\stackrel{>}{\scriptstyle\sim}}
\newcommand{\QED}{\mbox{\rule[-1.5pt]{6pt}{10pt}}}
\newtheorem{claim}{Claim}[section]
\newtheorem{theorem}[claim]{Theorem}
\newtheorem{proposition}[claim]{Proposition}
\newtheorem{corollary}[claim]{Corollary}
\newtheorem{lemma}[claim]{Lemma}
\newtheorem{conjecture}[claim]{Conjecture}
\newtheorem{remark}[claim]{Remark}
\newtheorem{remarks}[claim]{Remarks}

\title{Magnetic strip waveguides}
\date{}
\author{Pavel Exner$^{a,b}\!$, Hynek Kova\v{r}\'{\i}k$^{a,c}\!$}
\maketitle

\begin{quote}
{\small {\em a) Nuclear Physics Institute, Academy of Sciences,
25068 \v Re\v z \\ \phantom{a) }near Prague, Czechia
 \\ b) Doppler Institute, Czech Technical
University, B\v rehov{\'a} 7, \\ \phantom{a) }11519 Prague}
 \\ {\em c) Faculty of Mathematics and Physics, Charles
 University, \\
 \phantom{a) }V Hole\v{s}ovi\v{c}k\'{a}ch 2,18000 Prague;}\\
 \phantom{a) }\texttt{exner@ujf.cas.cz},
 \texttt{kovarik@ujf.cas.cz} }
\end{quote}

\begin{abstract}
\noindent We analyze the spectrum of the ``local" Iwatsuka model,
i.e. a two-dimensional charged particle interacting with a
magnetic field which is homogeneous outside a finite strip and
translationally invariant along it. We derive two new sufficient
conditions for absolute continuity of the spectrum. We also show
that in most cases the number of open spectral gaps of the model
is finite. To illustrate these results we investigate numerically
the situation when the field is zero in the strip being screened,
e.g., by a superconducting mask.
\end{abstract}


\section{Introduction}

There has been a renewed interest recently to problems of magnetic
transport in two-dimensional systems. Several papers
\cite{macr,fro1,biev} investigated the influence of a weak
disorder on current-carrying edge states in a halfplane
\cite{halp,macd}, or in a more general domain \cite{fro2}: roughly
speaking, it has been shown that away of the Landau levels the
transport survives for both the confining-potential and
Dirichlet-boundary border of the halfplane, and that the same is
true for Dirichlet regions containing an open wedge.

On the other hand, it is known for long that transport can be
achieved by a suitable variation of the magnetic field alone
\cite{iwa} if the latter is translationally invariant in one
direction. A simple example is represented by the field assuming
two different nonzero values in two halfplanes. It has an
illustrative semiclassical interpretation in terms of cyclotronic
radii \cite[Sec.~6.5]{cycon}. On the other hand, the spectrum of
this generic Iwatsuka model is purely absolutely continuous, while
classically there are many localized states; this shows that the
extended character of quantum states plays here an important role.

Furthermore, the transport does not require the existence of
different asymptotics; it is possible even if a variation of a
constant nonzero magnetic field is restricted to a planar strip,
i.e. its transverse profile has a compact support. Two sufficient
conditions for absolute continuity of the spectrum are known in
this case. In the original paper \cite{iwa} it was shown to be the
case if the field derivative has opposite signs to both sides of a
given point. Later Mantoiu and Purice \cite{mant} derived what
could be called the ``first and third quadrant" condition, i.e.
the requirement that $\pm B(x_1)\le \pm B(x_0)\le \pm B(x_2)$
holds for some $x_0$ and all $x_1\le x_0\le x_2$.

In the next section we discuss the ``local" Iwatsuka model. For
the sake of completeness we describe briefly the procedure of its
solution; we skip some details because the subject is discussed
thoroughly in \cite{iwa,mant}. We shall derive a pair of new
sufficient conditions for the absolute continuity. The first one
is a global assumption of the above type: it requires that the
field variation is nonzero and does not change sign in the strip
under consideration. The other is a rather weak {\em local}
condition: it is enough that the variation has at least a
powerlike growth from zero at the vicinity of the strip edge. The
main new ingredient in the second case is a semiclassical estimate
of the transverse eigenfunction tails. Notice also that we shall
not need the assumption of global field positivity employed in
\cite{iwa,mant}. Our conditions strengthen the existing results
and represent a step towards the proof of a conjecture put forth
in \cite[Sec.~6.5]{cycon} which states that {\em any} nonzero
(translationally invariant) field variation spreads the Landau
levels into a purely absolutely continuous spectrum.

One is also interested, of course, how the spectrum looks as a
set. In particular, it is natural to ask whether the spectral
bands can overlap, and if so, what is the number of open gaps
left. This question is addressed in Section~3. We show there that
the number of open gaps is finite provided the field variation has
a nonzero mean. The case of zero mean is more delicate and we
limit ourselves to demonstrating that the conclusion persists
under a particular assumption. We believe, however, that the
restrictions are rather of a technical nature and we conjecture
that the finiteness of open-gap number holds generally for the
local Iwatsuka model, up to possible regularity requirements
imposed on the field.

The above mentioned mathematical properties of the model represent
just one of possible motivations. On the physical side, there has
been an interest recently \cite{peet} to the behaviour of an
electron gas in a thin film under influence of a perpendicular
magnetic field which is screened in a part of the plane by a
superconducting ``mask". A circular ``anti-dot" represents a
solvable case which is treated in Ref.~\cite{peet} within several
different models.

Suppose now that the mask has the form of a straight
strip\footnote{A negative image situation with a field constant
within a strip and zero otherwise was treated in \cite{cal}. Here
the spectral picture is more complicated: there are transverse
eigenvalues of ``slow" enough states embedded in the continuum.
The same system and several related structures were studied also
in \cite{pm}.}. Then we get a particular case of our model, and
consequently, the magnetic field creates states which are
localized transversally and move along the strip. The form of
these states, in particular, their ``distance" from the strip is
controlled by the velocity; the transverse eigenenergies of fast
states are close to the Landau levels of the constant ``external"
field. In Section~4 we shall investigate numerically two variants
of such a system, one with an ideal screening and one with an
``overshoot": we will compute the band profiles and illustrate the
band overlapping.\footnote{After finishing this paper we learned
about recent results of Kim et. al. \cite{ring} who studied
properties of a nonmagnetic ring. In that case, of course, the
spectrum remains to be pure point but one can see a transport
analogous to the one discussed here in form of ``edge states"
moving along the ring.}


\setcounter{equation}{0} \section{Local Iwatsuka model}
\subsection{Description of the model}

We consider a charged particle confined to a plane and interacting
with a magnetic field perpendicular to the plane. Our basic
assumption is that field is translationally invariant in the
$y$-direction, nonzero and constant away of a strip of a width
$2a$: \\[2mm]
{\em (a)} the functional form of the field is
$$ B(x,y) = B(x) = B+b(x), \nonumber $$ 
where $B>0$ and $b$ is bounded and piecewise continuous with ${\rm
supp\,}b = [-a,a]$. With an abuse of notation, we employ the same
symbol for functions on $\mathbb{R}$ and $\mathbb{R}^2$ if they
are independent of one variable. \\[2mm]
As usual in translationally invariant situations we employ the
Landau gauge, i.e. we choose the corresponding vector potential in
the form
$$ A_x = 0, \quad A_y(x) =Bx+a(x) $$ \nonumber 
with
$$ a(x) := \int_0^xb(t)\,\D t $$ 
We adopt the natural system of units in which $ 2m = \hbar = c =
|e| = 1$ and assume also that the particle charge equals $-|e|$
having in mind an electron. Then the Hamiltonian $H\equiv H(B,b)$
of our system is
$$ H = ({\bf p}+{\bf A})^2 $$ 
with the appropriate domain in $L^2(\mathbb{R}^2)$.

Since $H$ commutes with $y$-translations, it allows for a standard
decomposition. Mimicking the argument of \cite[Sec.~2]{iwa} we
find that it is unitarily equivalent to the direct integral
\be \label{decomp} \int^{\oplus}H(p)\D p \ee
with the fiber space $L^2(\mathbb{R})$ and fiber operator
\be H(p) = -\pd_x^2 + (p+xB+a(x))^2 \label{fiber H} \ee
Since the function $a$ is bounded, the potential is for a fixed
$p\in\mathbb{R}$ dominated by the oscillator term, $D(H(p))=
D(-\pd_x^2)\cap D(x^2)$, the spectrum of $H(p)$ is purely discrete
and consists of a sequence of eigenvalues $\epsilon_n(p)$
accumulating at $+\infty$ --- see \cite[Thm XIII.16]{rees}.

In the absence of the perturbation $b$ the spectrum consists of
the Landau levels, $\{(2n+1)B:\:n\in \mathbb{N}_0\}$. We shall
first show that the latter belong to the spectrum in the perturbed
case too, at least as its accumulation points.
\begin{lemma} \label{limit Landau}
$\;\epsilon_n(p)\to (2n\!+\!1)B\,$ as $\,|p|\to\infty$ for any
$n\in\mathbb{N}_0$.
\end{lemma}
{\it Proof:}  We introduce a new variable $z$ and numbers
$a^{+},a^{-}$ by
\be z = x+p/B,\quad a^{\pm} = \int_0^{\pm a} b(x)\,\D x\,.
\label{zet}\ee
Then we have $H(p) = -\pd_z^2
 + V_p(z)$ with $ V_p(z) =
B^2(z+B^{-1}a(z-p/B))^2$. Outside $[p/B-a,p/B+a]$ the potential
$V_p(z)$ is quadratic and equal to $(z+a^{\mp})^2$ for $z <
(p/B-a)$ and $z
>( p/B+a)$,  respectively. Thus $V_p$ converges for $|p|\to\infty$
pointwise to the potential of the harmonic oscillator Hamiltonian,
\be H_0 = -\pd_z^2+V_0, \quad V_0(z) = Bz^2\,. \label{ho} \ee
Now we employ a simple trick \cite{bez}: we check the resolvent
convergence on functions $f = (H_0-\mu)\phi$ with $\phi\in
C_0^{\infty}(\mathbb{R})$. For any $\mu\in\rho(H(p))$ we have
$$ \left\|(H(p)-\mu)^{-1}f-(H_0-\mu)^{-1}f\right\| =
\left\|(H(p)-\mu)^{-1}(V_0\!-V_p)\phi\right\|\to 0 $$
as $|p|\to\infty$ in view of the compact support. However, such
$f$ form a dense set in $L^2(\mathbb{R})$, and therefore $H(p)\to
H_0$ in the strong resolvent sense and the claim follows by
Theorem VIII.14 of \cite{kato}. \quad \QED

\subsection{Absolute continuity}

Next we want to give sufficient conditions under which the
spectrum is absolutely continuous. We start with analyticity of
the perturbation.
\begin{lemma} \label{anal-A}
$\left\{H(p):\: p\in \mathbb{R}\cup \{\infty\}\right\}$ is an
analytic family of type (A). In particular, each
$\epsilon_n(\cdot)$ is an analytic function.
\end{lemma}
{\it Proof:} Let us first check the analyticity at infinity. We
write
$$ H(p) = H_0 + a(x)(2p+2Bx+a(x)) $$ 
where $H_0$ is oscillator Hamiltonian (\ref{ho}). Since $a$ is
bounded by assumption, we have to check $H_0$-boundedness of the
second term at the r.h.s. with a suitable relative bound. From the
Schwarz inequality we have
\be \|zf\| \leq \|(I+z^2)^{-1}z\| \|(I+z^2)f\| \leq
C(\|z^2f\|+\|f\|) \label{ineq1}\ee
since $z\mapsto z(1+z^2)^{-1}$ belongs to $L^2(\mathbb{R})$. Given
$f\in D(z^m)$ denote $f_{\eta}(z): = \eta^{3/2}f(\eta z)$; then
$\|z^mf_{\eta}\| = \eta^{1-m}\|z^mf\|$. Thus (\ref{ineq1})
acquires the form
 \be
\|zf\| \leq C\eta^{-1}\|z^2f\|+ C\eta\|f\|\,; \label{zf} \ee
in other words for any $\alpha'>0$ we can find $\beta'>0$ such
that
\be
\|zf\| \leq \alpha'\|z^2f\|+\beta'\|f\|. \label{ineq2}\ee
On the other hand,
\begin{eqnarray*} \|(P^2+B^2z^2)f\|^2 &\!=\!&
((P^4+ B^4z^4+ 2B^2Pz^2P+ B^2[P,[P,z^2]])f,f) \\ &\!=\!&
\|P^2f\|^2+ B^4 \|z^2f\|^2+ 2B^2\|zPf\|^2 -2B^2\|f\|^2
\end{eqnarray*}
with $P:=-i\pd_x$ holds for all $f\in \mathcal{S}(\mathbb{R})$,
which yields the following inequality:
\be \|z^2f\| \leq B^{-2}\|(P^2+ B^2z^2)f\| +\sqrt{2}B^{-1}\|f\|.
\label{ineq3} \ee
Combining (\ref{ineq2}) and (\ref{ineq3}) we arrive at
\be \|zf\| \leq \alpha\|(P^2+ B^2z^2)f\| +\beta\|f\| \label{rb1}
\ee
where $\alpha$ can be made arbitrarily small and $\beta =
\beta'+\alpha' B^{-1}\sqrt{2}$. To check the analyticity at any
point $p$ we employ the inequalities
\begin{eqnarray*} \|H_0f\| &\!\le\!& \|H(p)f\| +
\|a(\cdot)(2p+2B\cdot +a(\cdot))f(\cdot)\| \\ &\!\le\!&  \|H(p)f\|
+ 2B\|a\|_{\infty} \|zf\| + \|a\|_{\infty}^2 \|f\|.
\end{eqnarray*}
Applying (\ref{zf}) to the second term we get
$$ (1-2\alpha'B\|a\|_{\infty})  \|H_0f\| \le \|H(p)f\| + (2\beta'B
+ \|a\|_{\infty})\|a\|_{\infty} \|f\|, $$
and since $\alpha'$ can be made arbitrarily small, the sought
$H(p)$-boundedness of $H(p')\!-\!H(p)$ follows from (\ref{rb1}).
To conclude the proof, it is sufficient to apply \cite{kato}, Thms
VII.2.6 and VII.1.8. \quad \QED \\[4mm] \phantom{\quad}
Let $\psi_n(\cdot,p)$ denote the eigenfunctions of the operator
(\ref{fiber H}), i.e.
\be H(p)\psi_n(x,p) = \epsilon_n(p)\psi_n(x,p). \label{schr} \ee
Recall that without loss of generality $\psi_n$ can be chosen
real-valued. Following \cite{iwa} we put
\be Q_{n,p}(x) = (p+Bx+a(x))^2-\epsilon_n(p) \label{qnp} \ee
so (\ref{schr}) reads $\psi_n^{"}(x,p) = Q_{n,p}(x)\psi_n(x,p)$,
and
$$ l_n(x,p) = \psi'_n(x,p)^2-Q_{n,p}\psi_n(x,p)^2. $$ 
\begin{lemma} \label{iwa-ef}
{\rm (\cite[Lemma 3.4]{iwa})} $\;l_n(x,p)\to 0\;$ as
$\;|x|\to\infty$.
\end{lemma}
Finally, we have to know that the function
$$ f_n(x,p) := (p+xB+a(x))\psi_n(x,p)^2 $$
which determines the potential part of the energy form decays fast
enough for large $p$.
\begin{lemma} \label{decay}
For any $p$ large enough there is $c(p)>0$ such that
\be 5c(p)\, \e^{-p(x-x_0)} \ge f_n(x,p) \ge {c(p)\over 7}\,
\e^{-3p(x-x_0)} \ee
holds for all $-a\le x_0\le x \le a$.
\end{lemma}
{\it Proof:} By Lemma~\ref{limit Landau} there is $p_0>0$ such
that $Q_{n,p}(x) > 0$ holds for all $ x\in [-a,a]$ and $|p|>p_0$.
More than that, $Q_{n,p}(x)$ grows for a fixed $x$ as
$|p|\to\infty$, which makes it possible to to employ a
semiclassical form for the tails of the eigenfunctions: for large
enough $p$ we have
\be \label{sc} \psi_n(x,p) = {c_1(p)\over Q_{n,p}^{1/4}(x)}
\exp\left\{ -\int_{x_0}^x \sqrt{Q_{n,p}(\xi)}\, \D\xi \right\}
\left( 1+q_{n,p}(x)\right),\ee
where
\be |q_{n,p}(x)| \leq \exp\left [\frac{1}{2}\int_{x_0}^x
|F'_{n,p}(x')|\D x'\right ] -1 \label{err} \ee
and
$$ F_{n,p}(x) = \int\left\{Q_{n,p}^{-1/4}(x)\frac{\D^2}{\D
x^2}\left (Q_{n,p}^{-1/4}(x)\right )\right\}\D x $$
by \cite[Thm VI.2.1]{olv}. Substituting for $Q_{n,p}(x)$ from
(\ref{qnp}), we get
\begin{eqnarray*}
F'_{n,p}(x) &\!=\!& \frac{5}{4}\frac{(p+xB+a(x))
^2(B+b(x))^2}{\left [(p+xB+a(x))^2-\epsilon_n(p)\right ]^{5/2}}
\\ &\!-\!&
\frac{1}{2}\frac{(B+b(x))^2+b'(x)(p+xB+a(x))}{\left
[(p+xB+a(x))^2-\epsilon_n(p)\right ]^{3/2}} \end{eqnarray*}
Thus the integrand in (\ref{err}) can be made arbitrarily small
for large enough $p$. Consequently, to a fixed $\lambda > 1$ we
can always find $p_{\lambda}$ such that
\be \exp\left [\frac{1}{2}\int_{x_0}^x|F'_{n,p}(x')|\D x'\right ]
\leq \lambda. \label{lambda} \ee
holds for all $p>p_{\lambda}$. The representation (\ref{sc}) is
valid at the halfline $x\ge x_0$, so the coefficient $c_1(p)$ is
nonzero; without loss of generality we may suppose that it is
positive. The behaviour of $\psi_n(x,p)/\psi_n(x_0,p)$ is for
$x,x_0\in[-a,a]$ determined essentially by the exponential factor,
because $\left(Q_{n,p}(x_0)/Q_{n,p}(x) \right)^{1/2}$ can be then
included into the error term. Since $a(\cdot)$ is bounded and we
consider $x,x_0$ with a limited range, one has
$$ {1\over 2}p \le \sqrt{Q_{n,p}(\xi)} \le {3\over 2}p $$
for all $p$ larger than some $p_1>0$, and therefore
$$ \frac{9 c_1(p)^2}{2p}\, \e^{-p(x-x_0)} \ge \psi_n(x,p)^2 \ge
\frac{c_1(p)^2}{6p}\, \e^{-3p(x-x_0)} $$
if $p>\max(p_1,p_{3/2})$, where $p_{3/2}$ refers to $\lambda=3/2$
in (\ref{lambda}). To conclude the proof, notice that
$\lim_{p\to\infty} (p+xB+a(x))p^{-1} =1$ holds for $x\in [-a,a]$
and put $c(p):= c_1(p)^2$. \quad \QED
\begin{remarks} \label{mirror}
{\rm (i) For the sake of simplicity, we use large enough $p$ in
the above claim. However, the assumption {\em (a)} is not
sensitive to mirror transformation which changes $p$ to $-p$, so
the analogous result is valid as $p\to -\infty$. \\ (ii) We need
both an upper and a lower bound, hence we cannot employ
\cite{agmon, helf} as the authors of Ref.\cite{biev} did; we have
to resort to a more traditional semiclassical method.}
\end{remarks}
Now we are in position to prove the announced result under one of
the following additional assumptions: \\[2mm]
{\em (b)} $\;b(\cdot)$ is nonzero and does not change sign in
$[-a,a]$, \\ [2mm]
{\em (c)} let $\;a_{\ell}<a_r$, where we have put $a_{\ell}:=
\sup\{x:\: b(x)=0 \;{\rm in}\;(-\infty,x)\}$ and $a_r:= \inf\{x:\:
b(x)=0 \;{\rm in}\;(x,\infty)\}$. There exist $c_0,\delta>0$ and
$m\in\mathbb{N}$ such that one of the following conditions holds:
\be \label{local left} |b(x)|\ge c_0(x-a_{\ell})^m,\quad
x\in[a_{\ell},a_{\ell}+\delta) \ee
or
\be \label{local right} |b(x)|\ge c_0(a_r-x)^m,\quad
x\in(a_r-\delta,a_r]. \ee
\vspace{2mm}
\begin{theorem} \label{ac}
Assume (a) and (b), or (a) and (c); then $|\epsilon'_n(p)|>0$ for
each $n\in\mathbb{N}_0$ and all $|p|$ large enough. In particular,
the spectrum of $H$ is absolutely continuous.
\end{theorem}
{\it Proof:} To prove the absolute continuity, it is sufficient by
Thm XIII.86 of \cite{rees} to show that $\epsilon_n(\cdot)$ is not
constant for any $n\in\mathbb{N}_0$. The Feynman-Hellman formula
implies
\be \epsilon'_n(p) = \left(\psi_n(x,p),\frac{\D H(p)}{\D
p}\psi_n(x,p)\right ) =
2\int_{-\infty}^{\infty}(p+Bx+a(x))\psi_n(x,p)^2\D x \ee
\label{eps}
Let us first investigate the integral on the semi-infinite
intervals $(-\infty,-a]$ and $[a,\infty)$. Since
\be l'_n(x,p) = -2(B+b(x))(p+Bx+a(x))\psi_n(x,p)^2 \label{ln'} \ee
and $b(x)=0$ for $|x|>a$, we can write
\begin{eqnarray*}
\lefteqn{ 2 \int_{(-\infty,-a]\cup
[a,\infty)}(p+Bx+a(x))\psi_n(x,p)^2\,\D x } \nonumber \\ && =
-\frac{1}{B} \int_{(-\infty,-a]\cup [a,\infty)}l'_n(x,p)\,\D x =
\frac{1}{B} \Big[l_n(a,p)-l_n(-a,p)\Big] \end{eqnarray*}
where we have employed Lemma~\ref{iwa-ef}. Using (\ref{ln'}) for
the second time, we can rewrite the last expression further as
$$ -\frac{2}{B}\int_{-a}^a (B+b(x))(p+Bx+a(x))\psi_n(x,p)^2\,\D x
$$
and thus Eq.(\ref{eps}) acquires the form
\begin{eqnarray}
\epsilon'_n(p) &\!=\!& -\frac{2}{B}\int_{-a}^a
b(x)(p+Bx+a(x))\psi_n(x,p)^2\,\D x \nonumber \\ &\!=\!&
-\frac{2}{B}\int_{-a}^a b(x)f_n(x,p)\,\D x. \label{eps'}
\end{eqnarray}
The first claim of the theorem follows immediately since
$f_n(x,p)$ has a definite sign in $[-a,a]$ for $|p|$ large enough.

Assume now that {\em (c)} is valid. We shall consider the
condition (\ref{local left}) and suppose that $\,b(x)>0$ in
$(a_{\ell},a_{\ell}+\delta)$; the other cases can be treated in
the same way (cf. Remark~\ref{mirror}). In view of
Lemma~\ref{decay} the r.h.s. of (\ref{eps'}) can be estimated as
follows:
\begin{eqnarray*}
\lefteqn{\int_{a_{\ell}}^{a_{\ell}+\delta} b(x)f_n(x,p)\,\D x +
\int_{a_{\ell}+\delta}^{a_r} b(x)f_n(x,p)\,\D x} \\
&& \ge {c(p)\over 7}\, \int_{a_{\ell}}^{a_{\ell}+\delta} b(x)\,
\e^{-3p(x-a_{\ell})}\,\D x - 5c(p) \int_{a_{\ell}+\delta}^{a_r}
|b(x)|\, \e^{-p(x-a_{\ell})}\,\D x
\\
&& \ge {1\over 7}\, c_0 c(p)\, \int_{0}^{\delta} \xi^m\,
\e^{-3p\xi}\,\D\xi - 10\, ac_0 c(p)\, \|b\|_{\infty} \,
\e^{-p\delta}.
\end{eqnarray*}
Estimating the exponential function in the first integral at the
r.h.s. from below by $\max\left\{0, 1\!-\!3p\xi\right\}$ we get
\be \label{nonconst} \epsilon'_n(p) < -\,{2c_0 c(p) \over B}\,
\left\{ {(3p)^{-m-1} \over 7(m+1)(m+2)} - 10\, a\|b\|_{\infty} \,
\e^{-p\delta} \right\} < 0 \ee
for all sufficiently large $p$. \quad \QED
\begin{remark} \label{inf-anal}
{\rm Since the perturbation is analytic at infinity by
Lemma~\ref{anal-A} we know in fact more than (\ref{nonconst}):
under the assumptions of the theorem there are nonzero $c_{\pm}$
and positive integers $m^{(\pm)}$ such that
$$ 
\epsilon_n(p) = (2n\!+\!1)B + c_{\pm} p^{-m^{(\pm})} +
\mathcal{O}(p^{-m^{(\pm)}-1}) $$
as $p\to\pm\infty$.}
\end{remark}
%

\setcounter{equation}{0} \section{Number of gaps} \label{gaps}

In addition to the absolute continuity given by Theorem~\ref{ac}
we want to know how the spectrum of $H$ looks like as a set. It
follows from (\ref{decomp}) and (\ref{fiber H}) that $\sigma(H)$
consists of a union of spectral bands $I_n$:
$$ I_n = \Big\lbrack\, \inf_{p\in\mathbb{R}}\,\epsilon_n(p),\,
\sup_{p\in\mathbb{R}}\,\epsilon_n(p)\, \Big\rbrack\;; $$ 
the question is how many gaps between them remain open. In this
section we will show that their number is finite because the bands
overlap at sufficiently high energies.

\subsection{Field variation of a nonzero mean} \label{gaps 1}

We shall distinguish two different cases depending on whether the
functional $A[b] := \int_{-a}^a b(x)\,\D x$ vanishes or not.
Suppose first that $A[b]$ is nonzero. In that case we have:
\begin{proposition} \label{quant states}
Assume $\int_{-a}^ab(x)\,\D x \neq 0$. Let $n(E,p)$ and $n_0(E)$
be the numbers of eigenstates of $H(p)$ and $H_0$, respectively,
with the eigenenergy smaller than $E$. Then for any
$m\in\mathbb{N}_0$ there exist $p_0$ and $E(m,p_0)$ such that
$$ \left(n_0(E)-n(E,p_0)\right)\,{\rm sgn\,}A[b] > m $$ 
holds for all $E>E(m,p_0)$.
\end{proposition}
{\it Proof:} The assumption $\int_{-a}^ab(x)\,\D x\neq 0$ is
equivalent to $a^{-} \neq a^{+}$. Suppose for definiteness that
$A[b]<0$, i.e. $a^{-}>a^{+}$. Since we are interested in the high
energy limit we may accept without loss of generality that the
field variation support lies in the classically allowed region,
$E>(p_0+xB+a(x))^2$ for any $x\in [-a,a]$, and to employ the
Bohr-Sommerfeld quantization condition: by \cite[Thm 7.5]{tit} we
obtain then
\be \label{BS for H} \pi n(E,p_0) =
\int_{x^\ell(E)}^{x^r(E)}\sqrt{E-(p_0+xB+a(x))^2}\,\D
x+\mathcal{O}(E^0) \ee
where the classical turning points
$$ x^{\ell}(E) = - \frac{\sqrt{E}+p_0-a^{-}}{B}\,,\quad \ x^r(E) =
\frac{\sqrt{E}-p_0-a^{+}}{B}\,, $$
satisfy by assumption the inequalities $x^\ell(E)<-a$ and
$x^r(E)>a$.

The idea is to compare (\ref{BS for H}) with the analogous
expression for $H_0$. Since the spectrum is not affected by a
shift of the potential, we change the variable in (\ref{ho}) as
follows, $z\to z+a^{-}/B$, and obtain
$$ \pi n_0(E) =
\int_{x^\ell(E)}^{x_0^r(E)}\sqrt{E-(p_0+xB+a^{-})^2}\,\D
x+\mathcal{O}(1)\,. $$
We have used the fact that by construction the left turning point
is the same for both potentials, whereas the right one is moved to
$$ x_0^r(E) = \frac{\sqrt{E}-p_0-a^{-}}{B} $$
Since $a^{-} \neq a^{+}$ we have $a<x_0^r(E)<x^r(E)$. Taking
further into account that the two potentials coincide to the left
of $-a$, we may write the sought difference as
\ba \lefteqn{\pi\left[n(E,p_0)-n_0(E)\right] =
\int_{-a}^a\bigg\{\sqrt{E-(p_0+xB+a(x))^2}} \nonumber \\ && -
\sqrt{E-(p_0+xB+a^{-})^2} \bigg\}\,\D x +
\int_a^{x_0^r(E)}\bigg\{\sqrt{E-(p_0+xB+a^{+})^2} \nonumber \\ &&
- \sqrt{E-(p_0+xB+a^{-})^2}\bigg\} \,\D x
+\int_{x_0^r(E)}^{x^r(E)}\sqrt{E-(p_0+xB+a^{+})^2}\,\D x \nonumber
\\ && + \mathcal{O}(E^0) \label{nos}\,, \ea
the last term being simply a positive number independent of $E$.

In the first term at the r.h.s. of (\ref{nos}) we integrate over a
fixed interval, hence the result is $\mathcal{O}(E^{-1/2})$ as
$E\to\infty$ and may be absorbed into the error term. Furthermore,
choosing $p_0\geq -a^{+}-aB$ we achieve that the integrand in the
second term is positive, hence we have
$$ \pi\left[n(E,p_0)-n_0(E)\right ] \geq
\int_{x_0^r(E)}^{x^r(E)}\sqrt{E-(p_0+xB+a^{+})^2}\,\D x +
\mathcal{O}(E^0)\,. $$
It remains to estimate the last integral. Since the function is
nonnegative, decreasing and vanishes only if $x = x^r(E)$, we take
any $\delta\in(0,x^r(E)-x_0^r(E))$ and use a simple bound,
\ba \int_{x_0^r(E)}^{x^r(E)} \sqrt{E-(p_0+xB+a^{+})^2}\,\D x
&\!>\!& \int_{x_0^r(E)}^{x^r(E)-\delta}
\sqrt{E-(p_0+xB+a^{+})^2}\,\D x \nonumber \\  &\!\ge\!&
\sqrt{2B\delta\sqrt{E}-B^2\delta^2}\left
(\frac{a^{-}-a^{+}}{B}-\delta\right)\,, \nonumber \ea
which yields the sought result for $E$ large enough. The
inequality
$$ n(E,p_0)<n_0(E)-m $$
for $A[b]>0$ is obtained in the same way. \quad \QED
\begin{corollary}
If $A[b]\ne 0$ the number of open gaps in the spectrum of $H$ is
finite.
\end{corollary}
{\it Proof:} Let again $a^{-}>a^{+}$. Since $\epsilon_n(p)\to
(2n+1)B$ as $p\to\infty$ for a fixed $n\in\mathbb{N}$ by
Lemma~\ref{limit Landau}, it is sufficient to find $\tilde n$ and
$\tilde{p}$ such that
\be
\epsilon_{n+1}(\tilde{p})<(2n+1)B \label{eps smaller} \ee
holds for all $n\ge\tilde{n}$. This follows immediately from
Proposition~\ref{quant states} with $m=2$. In the opposite case,
$a^{-}<a^{+}$, the inequality (\ref{eps smaller}) is replaced by
$\epsilon_{n-1}(\tilde p)>(2n+1)B$. \quad \QED

\subsection{The case of zero mean: an example}    \label{gaps 2}

If $A[b]=0$ the situation is more complicated since the two
potentials differ only in a subset of the interval $(-a,a)$. We
restrict ourselves to an example. As above, put $a_{\ell}:=
\sup\{x:\: b(x)=0 \;{\rm in}\;(-\infty,x)\}$, and suppose that
{\em there is a number $c\in (a_{\ell},a)$ such that $a(x)<0$
holds in $(a_{\ell},c)$ and $a(c)=0$ .}

Let us show that the above conclusion about the finite number of
gaps persists in this case. We employ again the Bohr-Sommerfeld
condition (\ref{BS for H}) choosing $E$ and $p$ in such a way that
$c$ will be the right turning point, $E=(p+Bc)^2$. It may happen,
of course, that there is another classically allowed region to the
right of $c$ but changing the potential to $E$ there certainly
does not increase the number of bound states, i.e.
$$ \label{BS for H, zero mean} \pi n(E,p) \ge
\int_{x^\ell(E)}^{c}\sqrt{E-(p+xB+a(x))^2}\,\D
x+\mathcal{O}(E^0)\,. $$
This has to be compared with the number of oscillator states
corresponding to $a=0$. Introducing the variable $y:=c-x$ we get
\ba \pi [n(E,p) -n_0(E,p)] &\!\ge\!& \int_0^{c-a_{\ell}}
\sqrt{2\sqrt{E} -By +a(c\!-\!y)} \,\sqrt{By -a(c\!-\!y)} \,\D y
\nonumber
\\ &\!-\!& \int_0^{c-a_{\ell}} \,\sqrt{2\sqrt{E} -By} \sqrt{By} \,\D y
+\mathcal{O}(E^0)\,. \nonumber \ea
The difference of the integrands can be written as
\ba \lefteqn{ \sqrt[4]{4E} \left\lbrack \sqrt{1-{By
-a(c\!-\!y)\over 2\sqrt{E}}} \,\sqrt{By -a(c\!-\!y)} - \sqrt{1-{By
\over 2\sqrt{E}}} \,\sqrt{By} \right\rbrack } \nonumber \\ &&
=\sqrt[4]{4E} \left( \sqrt{By -a(c\!-\!y)} - \sqrt{By} \right)
+\mathcal{O}(E^{-1/4})\,. \phantom{AAAAAA} \nonumber \ea
The error term can be absorbed into that of (\ref{BS for H, zero
mean}), hence we get
$$ \pi [n(E,p) -n_0(E,p)] \ge \sqrt[4]{4E}\, \int_0^{c-a_{\ell}}
\left( \sqrt{By -a(c\!-\!y)} - \sqrt{By} \right) \D y
+\mathcal{O}(E^0)\,, $$
which gives the desired result since the integral is positive by
assumption.

The result can be modified for the other end of ${\rm supp\,}b$,
and in a similar vein one can treat cases where $a(x)$ is locally
positive close to an endpoint. We will not pursue the line further
since we believe that a deeper analysis of the behaviour around
the classical turning points would be useful, and at the same
time, that the property in question is valid generally:
\begin{conjecture}
The number of open gaps in the local Iwatsuka model is finite for
any nonzero $b$ satisfying minimal regularity assumptions such as
those formulated in Section.~2.1.
\end{conjecture}
%


\setcounter{equation}{0} \section{Example: a screened strip}

In this section we shall discuss in more detail the example
mentioned in the introduction. Apart of the physical interest
mentioned there, the system represents a solvable case which
allows us to illustrate the results of the previous sections.

\subsection{The ideal situation} \label{model1}

We shall consider first the simplest situation (without
``overshoot" in the terminology of \cite{peet}) when the distance
between the film supporting the electrons and the superconducting
strip is negligible and the magnetic field is perfectly screened
for $|x|<a$. Consequently, the field perturbation $b(x)$ is of the
form
\be b(x) = -B\Theta(|x|-a) \label{model 1} \ee
with the vector potential
\be A_x = 0,\quad A_y(x) = B(x-a)\Theta(x-a)+B(x+a)\Theta(-x-a)\,.
\ee
The fiber operator in the decomposition (\ref{decomp}) then looks
as follows
\be H(p) = -\pd_x^2+\left
[p+B(x-a)\Theta(x-a)+B(x+a)\Theta(-x-a)\right ]^2. \label{fiber
operator}  \ee
The real line decomposes into a union of several parts in which
the corresponding Schr\"odinger equation can be solved. Inside the
strip the potential term in (\ref{fiber operator}) is constant and
equal to $p^2$, so the eigenfunctions $\psi_n(x,p)$ of $H(p)$ are
there of the form
$$ \psi_n(x,p) = A\exp\left [\kappa_n(p)\, x\right ]+ B\exp\left
[-\kappa_n(p)\, x\right],\quad x\in[-a,a]\,, $$
where $\kappa_n(p):= \sqrt{p^2-\epsilon_n(p)}$. On the other hand,
for $x\notin [-a,a]$ the function $a$ is constant and equal to its
boundary values; in view of (\ref{zet}) they are
$$ a^{-} = -a^{+} = aB\,. $$
In other words, outside the strip the potential has a parabolic
shape and its two branches are shifted mutually by $2a$. Thus we
use the substitutions
\be
z = \left\{\begin{array}{ccc}
             x+a+p/B & \quad \dots \quad & x<-a \\
             x-a+p/B & \quad \dots \quad & x>a
                             \end{array} \right.
\label{subst} \ee
denote $r = Bz^2$ and look for solutions of Eq.(\ref{schr}) in the
form
\be \psi_n(z,p) = \exp(-r/2)\, u_n(r,p) \label{schr3} \ee
In a usual way we check that $u_n(r,p)$ satisfies the confluent
hypergeometric equation \cite{abr},
$$ r\, u''_n(r,p)+\left({1\over 2}-r\right)\, u'_n(r,p)-\alpha\,
u_n(r,p) = 0 $$ 
with $\alpha :=\frac{B-\epsilon_n(p)}{4B}$. Consequently, the
general solution in the ``parabolic regions" is of the form
$$ \psi_n(x,p) = \e^{-r/2}\left\{C_1\, M\left(\alpha,{1\over
2},r\right) +C_2\, U\left(\alpha,{1\over 2},r\right) \right\}\,.
$$ 
We should not forget, however, that $r$ in this formula comes from
two different substitutions (\ref{subst}), and moreover, that the
map $z\mapsto r$ is not a bijection. With the exception of the
case $p=0$ the real line therefore decomposes into four intervals.
In the outer regions the requirement of $L^2$ integrability leaves
the term with the $U$ function only, while in the middle
``parabolic regions" both functions are generally contained in the
linear combination. Denote
$$ x_0 = -\frac{p}{B}-a\, {\rm sgn}(p)\,. $$
For $p<0$ this point where the potential reaches zero is to the
right of the strip, $a<x_0$, and we have
\be
 \psi_n(x,p) = \left\{\begin{array}{lr}
               D_1\,\e^{-r/2}U(\alpha,\beta,r) &  -\infty<x<-a \\
                                 &   \\
              A\exp\left [\kappa_n(p)\, x\right ]
              +B\exp\left [-\kappa_n(p)\, x\right ] &
              -a\leq x\leq a \\
                                 &         \\
              \e^{-r/2}\left\{C_1\, M(\alpha,\beta,r)+C_2\,
              U(\alpha,\beta,r)\right\} &   a<x<x_0  \\
                                 &                         \\
              D_2\,\e^{-r/2}U(\alpha,\beta,r) &  x_0\leq x<\infty
               \end{array} \right.
\label{soln} \ee
and a similar expression for $p>0$ when $x_0<-a$. In the case when
$p=0$ we set $x_0 = -a$ by definition, then $C_1 = 0$ and $C_2 =
D_1$. Matching the solutions (\ref{soln}) smoothly at the points
$-a,\, a,\, x_0$ we arrive at a homogeneous system of six linear
equations for the coefficients $ A,B,C_1,C_2,D_1$, and $D_2$,
which yields an implicit equation for the energy levels
$\epsilon_n(p)$.

\subsection{The model with an overshoot}

The above model is, of course, idealized because in reality the
magnetic field is never fully screened and the vector potential
changes sharply near the edges of the screening strip. Another
simplified description takes this effect into account by putting

\be b(x) = -B\Theta (a-|x|)+ B|x|\delta(|x|-a) \label{model 2} \,, \ee

which corresponds the vector potential
$$ A(x) = xB\Theta (|x|-a)\,. $$
Recall that a numerical analysis performed in \cite{peet} for a
screening by a disk suggests that a realistic field profile should
lie between these two extrema. The fiber operator in
(\ref{decomp}) now reads
\be H(p) = -\pd_x^2+[p+Bx \Theta (|x|-a)]^2. \label{Hp} \ee
Inside the strip we get the same solution as in the model without
an overshoot. However, the boundary values of the function $a(x)$
given by (\ref{zet}) are now equal to each other,
$$ a^{\pm} = \lim_{\eta\to 1+} \int_0^{\pm \eta a} b(x)\,\D x =
0\,, $$
hence for $x\notin [-a,a]$ we use the same substitutions to the
left and to the right of the strip,
$$ z' = x +p/B \quad \mathrm{and} \quad  r'= B(z')^2, $$
In analogy with Sec.~\ref{model1} we find the overshoot solutions
$\tilde{\psi}_n(x,p)$ of (\ref{schr3}) in the form
\be
 \tilde{\psi}_n(x,p) = \left\{\begin{array}{lr}
                          D'_1\,\e^{-r'/2}U(\alpha,\beta,r') &  -\infty<x<-a \\
                                     &   \\
                            A'\exp\left [\kappa_n(p)\, x\right ]
                            +B'\exp\left [-\kappa_n(p)\, x\right ] &
                            -a\leq x\leq a \\
                                 &         \\
                     \e^{-r'/2}\left\{C'_1\, M(\alpha,\beta,r')+C'_2\,
                   U(\alpha,\beta,r')\right\} &   a<x<x'_0  \\
                                 &                         \\
                  D'_2\,\e^{-r'/2}U(\alpha,\beta,r') &  x'_0\leq x<\infty
                                                 \end{array}
                                                 \right.
                                                 \phantom{AAA}
\label{soln2} \ee
where $x'_0 = -p/B$. Matching the functions (\ref{soln2}) smoothly
at the points $-a,\, a,\, x_0$ we arrive again at an implicit
equation for the energy levels $\tilde{\epsilon}_n(p)$.

\subsection{Numerical results}

\begin{figure}[!t]
\begin{center}
\includegraphics[height=12cm, width=\textwidth]{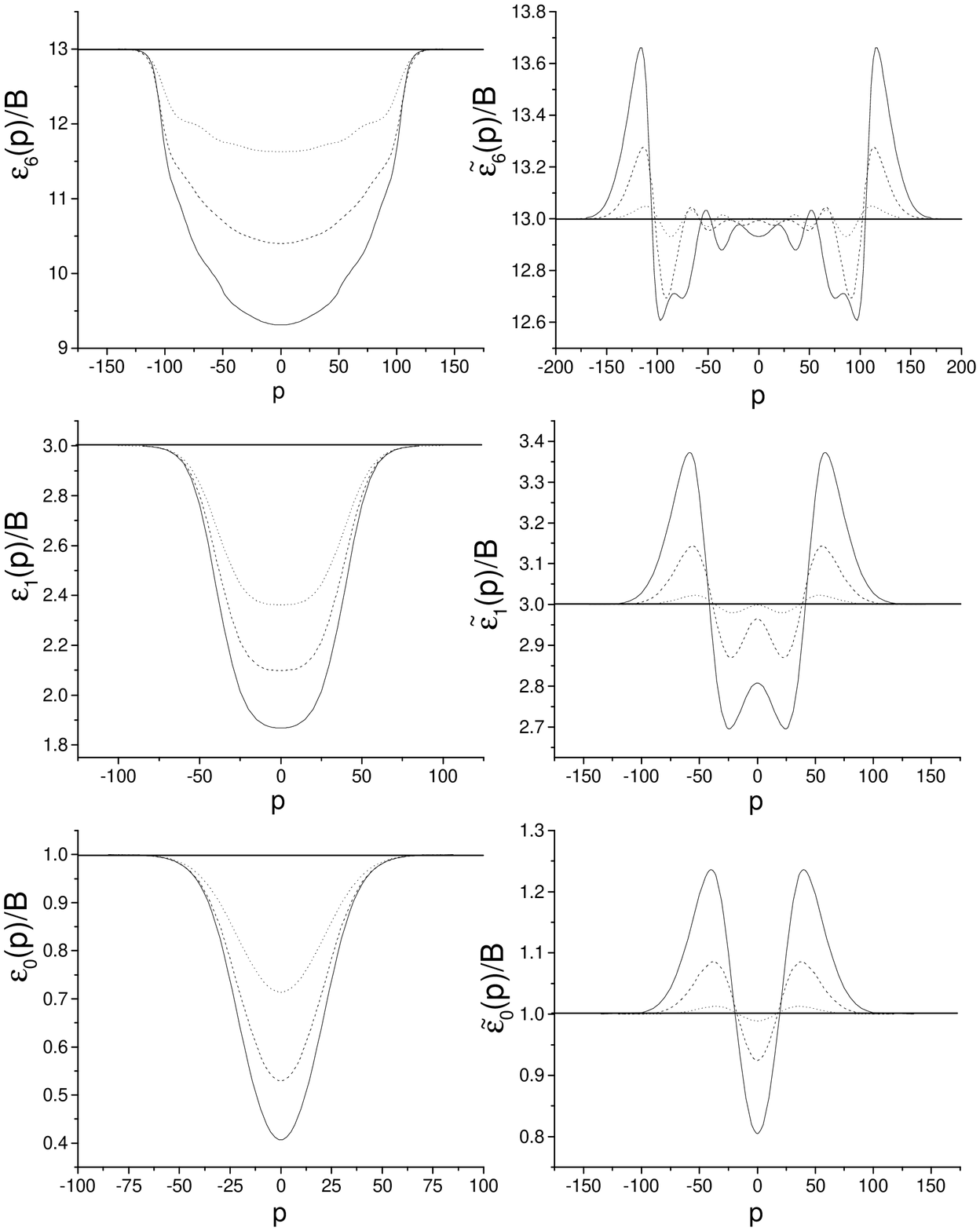}
\end{center}
\caption{The band functions in the units of $B$ for $B=10^3$,
$a=0.01$ (dotted line), $a=0.02$ (dashed line) and
$a=0.03$ (full line). (Left) the ideal situation. (Right) the model
with an overshoot. The thick lines show the Landau levels.}
\label{fig1}
\end{figure}

Using the Ans\"atze (\ref{soln}) and (\ref{soln2}) for the two
described models, we have solved the matching conditions
numerically for several values of the parameters. The obtained
band functions, denoted $\epsilon_n(p)$ and
$\tilde{\epsilon}_n(p)$ for the case without and with the
overshoot, respectively, are plotted in Fig.~\ref{fig1}.

\begin{figure}[!t]
\begin{center}
\includegraphics[angle=-90, width=\textwidth]{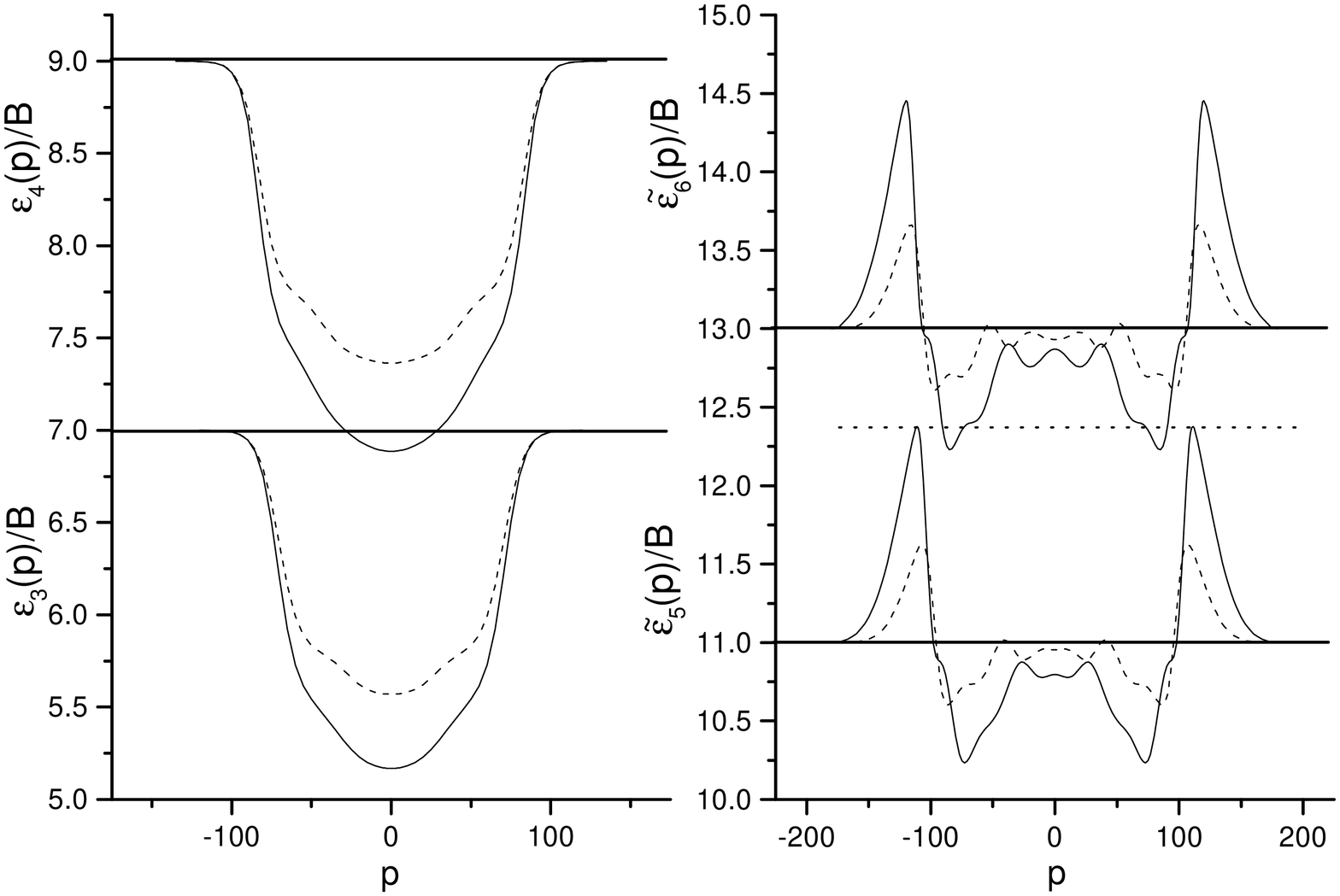}
\end{center}
\caption{Overlap of the spectral bands. (Left) the ideal
situation: $a = 0.015$ (dashed line), $a = 0.02$ (full line).
(Right) the model with an overshoot: $a = 0.03$ (dashed line), $a
= 0.045$ (full line). The field is always $B = 10^3$.}
\label{fig2}
\end{figure}

We see that in the ideal situation (without an overshoot) the band
functions $\epsilon_n(p)$ have the only stationary point at $p=0$.
This is true for any $n$ as it follows from Eq.~(\ref{eps'}) which
now acquires the form
\be \epsilon'_n(p) = 2p\int_{-a}^a\psi_n(x,p)^2\,\D x\,. \ee
The situation is completely different in the presence of an
overshoot. Here the fiber operator (\ref{Hp}) can be rewritten as
a sum of $H_0$ with the perturbation supported in $[-a,a]$.
Consequently, the functions $\tilde{\epsilon}_n(p)$ exhibit in
general many oscillations corresponding to the harmonic oscillator
solutions. Following this argument we would expect that in higher
levels these oscillations will be partially ``ironed" out for
sufficiently large $a$. As an illustration, see the curves for
$n=6$ in Figs.~\ref{fig1} and \ref{fig2}.

\par
In Section \ref{gaps 1} we have shown that if the functional
$A[b]$ is nonzero the number of open spectral gaps is finite.
Since the field variation (\ref{model 1}) of the ideal model gives
$A[b] = -2Ba$, it follows from Proposition \ref{quant states} that
for a large enough band index $n$, depending on $a$ and $B$, the
width of the spectral band $I_n$ exceeds $2B$, the distance
between Landau levels. Moreover, applying the Bohr-Sommerfeld
quantization rule we can easily derive an explicit condition under
which this happens:
\be n_\mathrm{overlap}(a,B) \ageq \frac{1}{2}+\frac{\pi^2}{8a^2B}
\label{overlap-c} \ee
This is shown in Fig \ref{fig2}. In the case of the overshoot
model with the field variation given by (\ref{model 2}) we have no
simple condition analogous to (\ref{overlap-c}). Nevertheless, the
assumptions of the example of Section \ref{gaps 2} are satisfied
here so we know that at sufficiently high energies the spectral
bands overlap in this case too.

On a heuristic level, this can be understood from the behavior of
the harmonic oscillator solutions. It is well known that for large
$n$ they are strongly oscillating but the squared modulus smoothed
over a small interval tends as $n\to\infty$ to the probability
density of finding the classical oscillator at a given point
\cite[Section~8.3]{beh}. The latter diverges at the turning
points. Taking for the difference between $\tilde{\epsilon}_n(p)$
and the corresponding Landau level the first-order perturbation
theory expression, we expect therefore this quantity to reach its
maximum for the values of $p$ corresponding to the vicinity of the
turning points. Fig~\ref{fig2} illustrates that it is indeed the
case.

\begin{figure}[!t]
\begin{center}
\includegraphics[width=\textwidth]{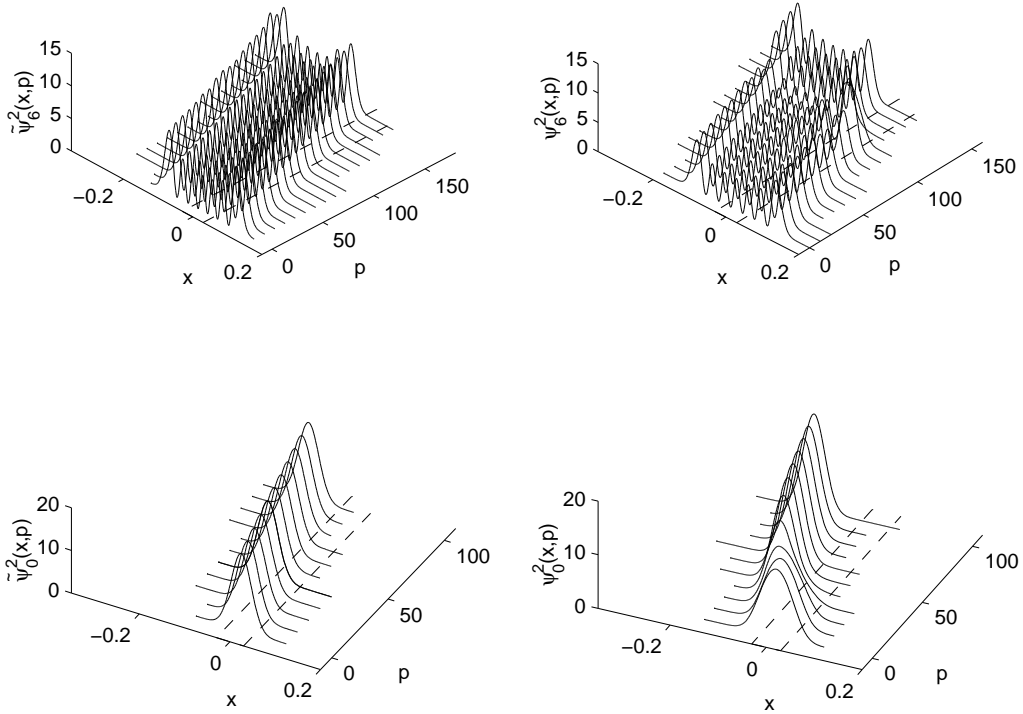}
\end{center}
\caption{The squared eigenfunctions for $n=0,\,6$,  and different
values of $p$. The dashed lines indicate the position of the
nonmagnetic strip with $a = 0.03$, the field is  $B = 10^3$.}
\label{fig3}
\end{figure}

\par
Once we have solved the matching conditions for the band functions
we can also find the  coefficients in (\ref{soln}) and
(\ref{soln2}) to obtain complete eigenfunctions $\psi_n(x,p)$ and
$\tilde{\psi}_n (x,p)$, respectively. Fig~\ref{fig3} shows the
corresponding probability densities for ground state and the
excited state with $n=6$. In the absence of an overshoot it can be
seen that as the screened strip $[-a,a]$ crosses the center of the
distribution , each particular peak of the density becomes broader
and more smeared. On the other hand, for large $|p|$ when the
``bulk" of the $|\psi_n(x,p)|^2$ support lies outside the strip,
its shape is close to that of the harmonic-oscillator probability
density. In the case of an overshoot the picture is similar but
the plot shows a significant deformation for the value of $|p|$
corresponding to the classical turning points.


\subsection*{Acknowledgment}

The research has been partially supported by GA AS under the
contract 1048801.



\begin{thebibliography}{99}

%
\bibitem[AS] {abr} M.~Abramowitz and I.A.~Stegun: {\em Handbook of Mathematical
Functions,} National Bureau of Standards 1964.\vspace{-1.8ex}
%
\bibitem[Ag]{agmon} S.~Agmon: {\em Lectures on exponential decay of
solutions of second-order elliptic equations,} Princeton
University Press 1982.\vspace{-1.8ex}
%
\bibitem[BEH]{beh} J.~Blank, P.~Exner, M.~Havl\'{\i}\v cek: {\em
Hilbert Space Operators in Quantum Physics,} AIP Press, New York
1994.\vspace{-1.8ex}
%
\bibitem[BEZ]{bez} F.~Bentosela, P.~Exner P., V.A.~Zagrebnov: Electron
trapping by a current vortex, {\it J.Phys. A: Math. Gen.} {\bf 31}
(1998), L305-311. \vspace{-1.8ex}
%
\bibitem[BP]{biev} S.~De Bi\`evre, J.V.~Pul\'{e}: Propagating edge states
for a magnetic Hamiltonian, {\it mp-arc 99-78}. \vspace{-1.8ex}
%
\bibitem[Ca]{cal} M.~Calvo: Exactly soluble two-dimensional electron gas
in magnetic-field barrier, {\it Phys. Rev.} {\bf B48} (1993),
2365-2369. \vspace{-1.8ex}
%
\bibitem[CFKS]{cycon} H.L.~Cycon, R.G.~Froese, W.~Kirsch, B.~Simon:
{\em Schr\"odinger Operators with Applications to Quantum
Mechanics and Global Geometry,} Springer, Berlin 1987.
\vspace{-1.8ex}
%
\bibitem[FGW1]{fro1} J.~Fr\"ohlich, G.M.~Graf, J.~Walcher: On the extended
nature of edge states of quantum Hall Hamiltonians, {\it
math-ph/9903014}. \vspace{-1.8ex}
%
\bibitem[FGW2]{fro2} J.~Fr\"ohlich, G.M.~Graf, J.~Walcher: Extended
quantum Hall edge states: general domains, {\it mp$\_$arc 99-327}.
\vspace{-1.8ex}
%
\bibitem[Ha]{halp} B.I.~Halperin: Quantized Hall conductance,
current carrying edge states, and the existence of extended states
in two-dimensional disordered potential, {\it Phys. Rev.} {\bf
B25} (1982), 2185-2190.\vspace{-1.8ex}
%
\bibitem[He]{helf} B.~Helffer: {\em Semiclassical Analysis for
the Schr\"odinger Operator and Applications,} Lecture Notes in
Math., vol.~1336, Springer, Berlin 1966.\vspace{-1.8ex}
%
%
\bibitem[Iw]{iwa} A.~Iwatsuka: Examples of absolutely continuous
Schr\"{o}dinger operators in magnetic fields, {\it Publ. RIMS}
{\bf 21} (1985), 385-401.\vspace{-1.8ex}
%
\bibitem[Ka]{kato} T.~Kato: {\em Perturbation Theory for Linear
Operators,} Springer, Heidelberg 1966.\vspace{-1.8ex}
%
\bibitem[KISC]{ring} N.~Kim, G.~Ihm, H.-S.~Sim, K.J.~Chang: Electronic
structure of a magnetic quantum ring, {\it Phys. Rev.} {\bf B60}
(1999), 8767-8772. \vspace{-1.8ex}
%
\bibitem[MS]{macd} A.H.~MacDonald, P.~St\v{r}eda: Quantized Hall effect
and edge currents, {\it Phys. Rev.} {\bf B29} (1984),
1616-1619.\vspace{-1.8ex}
%
\bibitem[MMP]{macr} N.~Macris, Ph.A.~Martin, J.V.~Pul\'{e}: On edge states
in semi-infinite quantum Hall systems, {\it cond-mat/9812367}.
\vspace{-1.8ex}
%
\bibitem[MP]{mant} M.~Mantoiu, R.~Purice: Some propagation properties
of the Iwatsuka model, {\it Commun. Math. Phys.} {\bf 188} (1997),
691-708. \vspace{-1.8ex}
%
\bibitem[Ol]{olv}
F.W.J.~Olver: {\em Asymptotics and Special Functions}, Academic
Press, New York 1974. \vspace{-1.8ex}
%
\bibitem[PM]{pm} F.M.~Peeters, A.~Matulis: Quantum structures created
by nonhomogeneous magnetic fields, {\it Phys. Rev.} {\bf B48}
(1993), 15166-15174. \vspace{-1.8ex}
%
\bibitem[RS]{rees} M.~Reed and B.~Simon: {\em Methods of Modern Mathematical
Physics, I. Functional Analysis, IV. Analysis of Operators,}
Academic Press, New York, 1972, 1978.\vspace{-1.8ex}
%
\bibitem[RPM]{peet} J.~Reijniers, F.M.~Peeters, and A.~Matulis: Quantum
states in a magnetic anti-dot, {\it Phys. Rev. B} {\bf 59} (1999),
2817-2823.\vspace{-1.8ex}
%
\bibitem[Ti]{tit} E.C.~Titchmarsh: {\em Eigenfunction Expansions,} Clarendon
Press, Oxford 1949.
%
\end{thebibliography}
\end{document}